# EAGLE ISS – a modular twin-channel integral-field near-IR spectrograph

Peter Hastings[a], Brian Stobie[a], Sébastien Vivès[b], Pascal Vola[b], Martyn Wells[a], Christopher Evans[a]

[a] UK Astronomy Technology Centre, Royal Observatory, Edinburgh, EH9 3HJ, Scotland

[b] Observatoire Astronomique de Marseille-Provence Laboratoire d'Astrophysique de Marseille, Pôle de l'Etoile Site de Château-Gombert 38, rue Frédéric Joliot-Curie 13388 Marseille, France

## ABSTRACT

The ISS (Integral-field Spectrograph System) has been designed as part of the EAGLE Phase A Instrument Study for the E-ELT. It consists of two input channels of 1.65x1.65 arcsec field-of-view, each reconfigured spatially by an image-slicing integral-field unit to feed a single near-IR spectrograph using cryogenic volume-phase-holographic gratings to disperse the image spectrally. A 4k x 4k array detector array records the dispersed images. The optical design employs anamorphic magnification, image slicing, VPH gratings scanned with a novel cryo-mechanism and a three-lens camera. The mechanical implementation features IFU optics in Zerodur, a modular bench structure and a number of high-precision cryo-mechanisms.

## 1. INTRODUCTION

EAGLE is a multi-object, AO-corrected, integral-field spectrograph designed for use at a gravity-invariant focal station of the E-ELT [1]. This paper describes the IFU and Spectrograph System (ISS) for EAGLE. Ten identical ISSs permit up to twenty science observations to be made simultaneously. Earlier units in the optical path of EAGLE collect and analyse light from natural- and laser-guide stars, and select, collect and correct [2] light from the science sources. Each ISS takes light from two independent science objects and splits it both spatially and then spectrally before focusing it on the detector array. The data from the array are re-formatted to form 3-D 'data cubes' that can be sliced in one direction to show the appearance of the entire target at a single wavelength, or at right angles to produce spectra across the target.

The optics, mechanisms and internal structure of the ISS are cooled to 180K with the detector cooled to 80K to reduce thermal background radiation on the detector to levels less than the background in the science path. There are two optical sub-systems within the ISS. The integral-field unit (IFU) splits the light spatially using an image-slicer and forms a continuous slit and common exit pupil for both input sources. A single spectrograph disperses the light by wavelength and images the resultant spectra from the two sources on two halves of a 4K x 4K detector. The volume available for each ISS is a cuboid, 700mm wide, 1150mm deep and 1800mm high. The total mass allowed for each ISS is 500kg. One of the challenges of the instrument is the need to design opto-mechanical units which can be entirely interchangeable. While routinely achieved in commercial mass-production it is relatively new to the 'small-batch' world of astronomical instrumentation.

## 2. INTEGRAL-FIELD UNIT

The IFU design is related to those previously used in a number of instruments, including UIST [3], KMOS [4], MIRI [5], MUSE [6], and a demonstrator of the SNAP spectrometer [7]. As in these examples, the IFU consists entirely of reflective optics, and is therefore achromatic. Each IFU consists of three main modules. The Anamorphic Optical Unit re-images the input field onto the slicing mirrors and introduces an anamorphic magnification of the field so that at the slicing mirror each spatial resolution element projects onto two pixels in the spectral direction to ensure correct (Nyquist) sampling in the dispersion direction while maintaining square pixels on the sky. It also re-images each entrance pupil to a reflective cold stop and brings the two science channels side-by-side at the slicer plane. The Image Slicer Unit re-arranges each original 2-D source at the input focal planes of the ISS to form half of an effective entrance slit for the Spectrograph. The Collimator Mirror Module produces the output collimated beam and re-images the pupil for all slices of both sources onto a common pupil at the dispersing element of the Spectrograph. An aluminium chassis acts as a mechanical foundation for the three modules, ensuring their mutual alignment and acting as a cooling path for the optics. The main requirements for one channel of the IFU are summarised below in Table 1.

Table 1: Main IFU requirements

| Field Size | 1.65 x 1.65 arcsec on sky |
|---|---|
| Spatial Sampling | 37.5 milliarcsec / pixel |
| Slice width (sub-slit) | 37.5 milliarcsec on sky |
| Number of slices | 44 slices each 44 pixels long |
| Input Beam | 30 mm diameter pupil image in collimated beam |
| Output Beam | Collimated with an elliptical pupil of 106mm (spatial) x 53mm (spectral) at the grating of the Spectrograph |
| Anamorphic Factor | 2 : 1 |

## 2.1 Anamorphic Optical Unit (AOU)

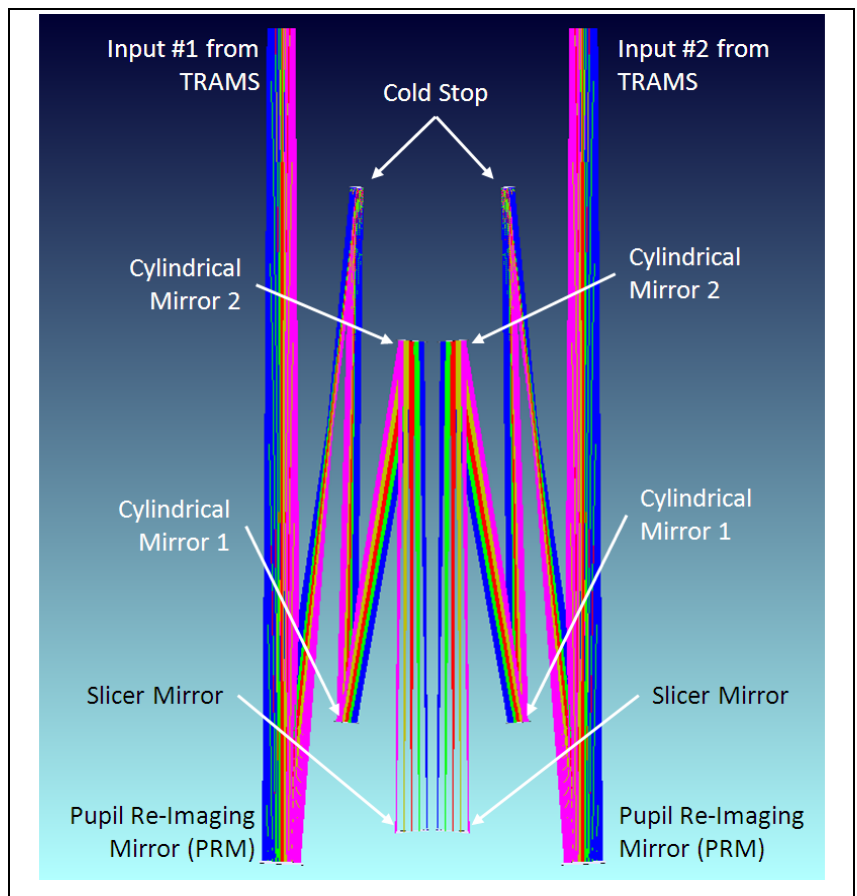

Figure 1: Optical layout of the two Anamorphic Optical Units

For the AOU shown in Figure 1 a single channel consists of four mirrors:

a. The Pupil Re-imaging Mirror (PRM) which forms the pupil image on the Cold Stop Mirror (CSM).

b. The Cold Stop which re-images the field of view on the slicer mirror with the correct sampling (F/130). This Cold Stop is the main Cold Stop of the overall EAGLE optical train.

c. Two cylindrical mirrors (CM1, CM2) which create the anamorphism in the spectral direction (F/260).

Two AOUs are mounted on the same optical chassis symmetrically (but chirally) to bring the target scientific fields side-by-side on adjacent arrays of image-slicing mirrors.

## 2.2 Image Slicer Unit (ISU)

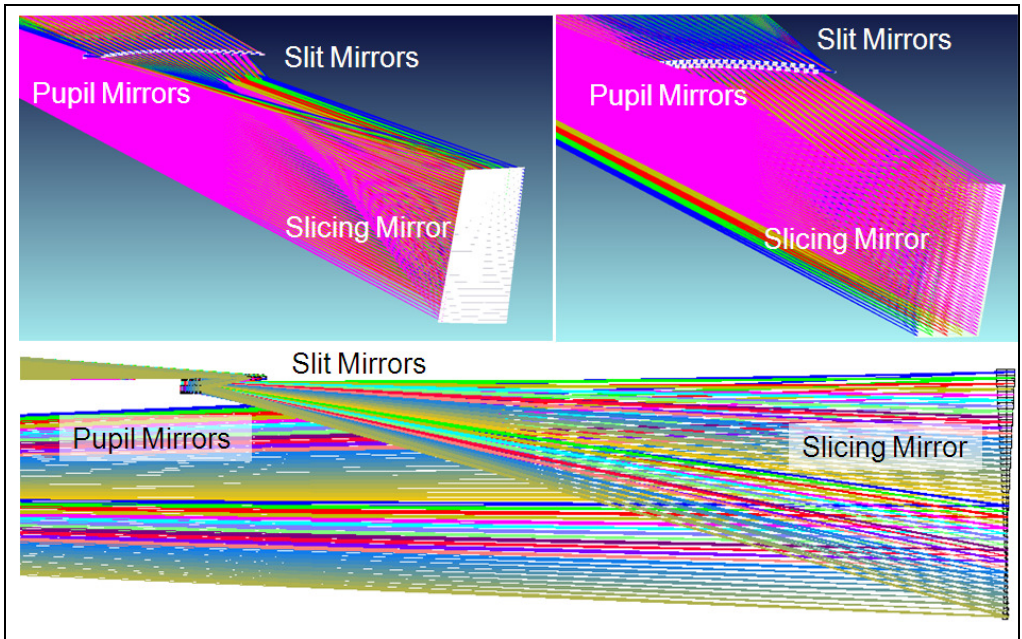

Figure 2: Optical layout of the Image Slicer Unit.

Figure 2 shows the ISU consisting of three optical assemblies: the two slicing mirror arrays at the output image planes of the AOU, an array of pupil mirrors and an array of slit mirrors.

The Image-Slicing Mirror Array consists of a stack of forty-four concave spherical mirrors (2 mm wide and 44 mm long) which slice the anamorphic field and produce an array of individual images of the cold stop (pupil) on the pupil mirrors. All slices have the same radius of curvature of about 522 mm. The number of slices defines the field-of-view in the spectral direction. The proposed method for the manufacture of the slicing mirrors is by glass polishing. Slices are made in Zerodur and assembled by optical contacting. The current design of the slicer (stack of forty-four slices) is compatible with an innovative method [8] enabling the manufacture of one or more stacks of slices by a single standard polishing process thus reducing both the time and cost of production.

The Pupil Mirror Array consists of two staggered lines of twenty-two rectangular mirrors each disposed along the spatial direction in a concave curve. Each pupil mirror re-images its own slice of the anamorphic field on to a dedicated slit mirror located at the input focal plane (slit plane) of the Spectrograph. The pupil mirrors are rectangular mirrors with an optical aperture of 3.4 mm x 2.6 mm. The elliptical sub-pupils are each decentred within the physical aperture of the mirror to a greater or lesser extent. All the pupil mirror surfaces are spherical and concave but have individual radii of curvature ranging from 27mm to 44mm.

The Slit Mirror Array consists of two staggered lines of twenty-two rectangular mirrors each with an optical aperture of 3.6 mm x 0.9 mm. Each slit mirror re-images the telescope pupil on its pupil mirror to the entrance pupil of the Spectrograph. The surface of each slit mirror is spherical and concave and has a radius of curvature of about 37 mm. Because slit mirrors are located at the output images of the sliced field, they are used to define the field of view of the system in the spatial (along slice) direction. If needed, additional masks made of blackened aluminium can be located there, to control stray and scattered light in the IFU. The manufacturing process for the pupil- and slit-mirrors is similar to that proposed for the slicing mirrors. For the pupil mirrors, with their various radii of curvature, diamond machining is a promising alternative that will continue to be investigated.

The Collimator Mirror Module is common to all slices of the two scientific channels in the ISS. It produces a collimated beam for each slice and places the pupil images at a common, elliptical, exit pupil. The collimator is a simple spherical mirror with a rectangular aperture of ~600 x 110mm and is readily made in glass. It is followed by a fold mirror which enables the ISS to fit in its allocated volume.

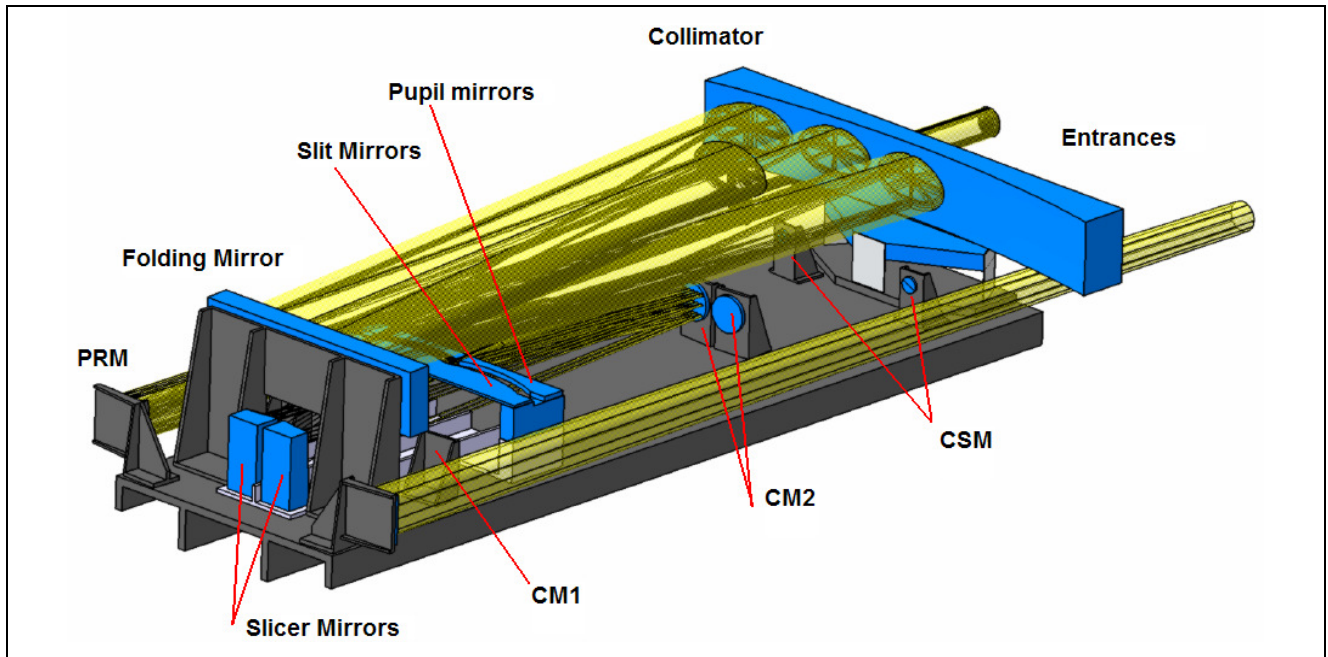

Figure 3: Mechanical details of the IFU

As shown in Figure 3, the IFU is packaged as a single unit that can be aligned and tested before its integration with the ISS. The IFU bench is a simple rectangular fabrication made from an aluminium alloy matching the thermal expansion of the Cryo-structure. In the current concept, the total mass of IFU is estimated around 40 kg, including 12 kg for optical components.

The main surface of the bench is parallel to, and 50 mm below, the plane containing the two incoming chief rays. The two sets of four mirrors forming the anamorphic optics are mounted on their own supports which are shimmed individually and fixed to the bench. The components of the slicing unit are treated as their own independent sub-assembly:

1- The arrays of forty-four slicer mirrors, pupil mirrors and slit mirrors are, from a mechanical point of view, treated as single optical parts.

2- The arrays must obviously be aligned to each other with particular attention since the forty-four optical paths must be adjusted together.

For these reasons, the combination of slicer, pupil and slit mirrors are mounted, aligned and tested as an independent, interchangeable, unit before integration with the rest of the IFU.

**2.3     Interchangeable Mounting**

Three machined locations on the IFU bench carry a double-shim mounting arrangement for ensuring interchangeability. One shim is attached permanently to the IFU, the other to a similar machined feature on the Cryo-structure. This principle is used throughout the ISS with the interface plane(s) between mating units being defined by the contacting surfaces of permanently-attached machined shims. For all units the relationship between the optical interface(s) and the physical interfaces is made identical. This allows the interchangeability of units between and within ISSs. The thermal conduction of the 'double shim' interface will be relatively low and will be supplemented by thermal straps as needed.

## 3. SPECTROGRAPH

The spectrograph for EAGLE is slightly unusual in that its reflective collimator forms part of the IFU module. It uses Volume-Phase Holographic Gratings (VPHGs) [9] and this choice determines the beam sizes, layout and camera design. The baseline detector is a 4K x 4K near-IR array with 15 x 15µm pixels. This images the two spectra from the IFU via a single set of spectrometer optics (scan mirrors, grating and camera). The camera is a refractive design containing three lenses in $CaF_2$, ZnS and ZnSe. Since the three lens camera is not achromatic, a Camera Focus Module allows the last two lenses and the detector to be moved together axially and for the detector to be tilted about an axis parallel to the slit. The main drivers/requirements for this design of the EAGLE Spectrograph are :

Four wavebands, namely: IZ, YJ, H, K

4K x 4K detector with 15x15 µm pixels

Optical interface matched to the IFU

Slit width = 37.5mas, slit length 2(IFUs) x 2048 x 37.5mas = 153.6 arcsec

Two pixels per slit width

Resolving powers of R = 4000 and R = 10000

### 3.1 Volume-Phase Holographic Gratings

VPHGs have high efficiency (>70%) and, when used in transmission in a Littrow configuration, do not introduce any anamorphic magnification when gratings are used with different angles of incidence. This has enabled more than one resolving power to be provided for the ISS in a compact manner. The sampling of the spectra (pixels/resolution element) can be kept constant over the near-IR bands for both the normal resolving power (R=4000) and high resolving power (R=10000) modes of EAGLE. Cryogenic characterisation of larger VPHGs is currently in progress at the UKATC.

### 3.2 Scan Mirrors

The Scan Mirrors (SMs) simplify the use of VPHGs by removing the need to articulate the camera [10] to match the output angle of the diffracted beam. The two mirrors are mounted in a single module and can be scanned in angle, using a single motor, over a range of some 15 degrees. They are tilted by equal angles, but opposite senses, keeping the beam from the IFU and the beam to the camera in a constant relationship. They allow different passbands to be selected from a single VPHG and allow the use of normal- and high-resolution VPHGs with no changes to the collimator or camera optics.

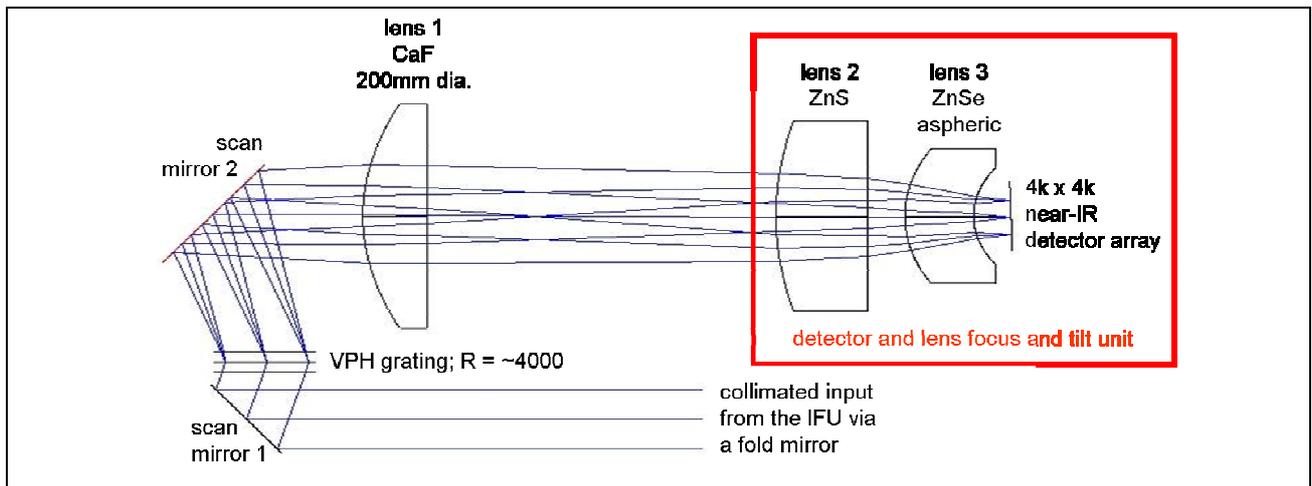

Figure 4: Layout of the Spectrograph optics for a spectral resolution of 4000

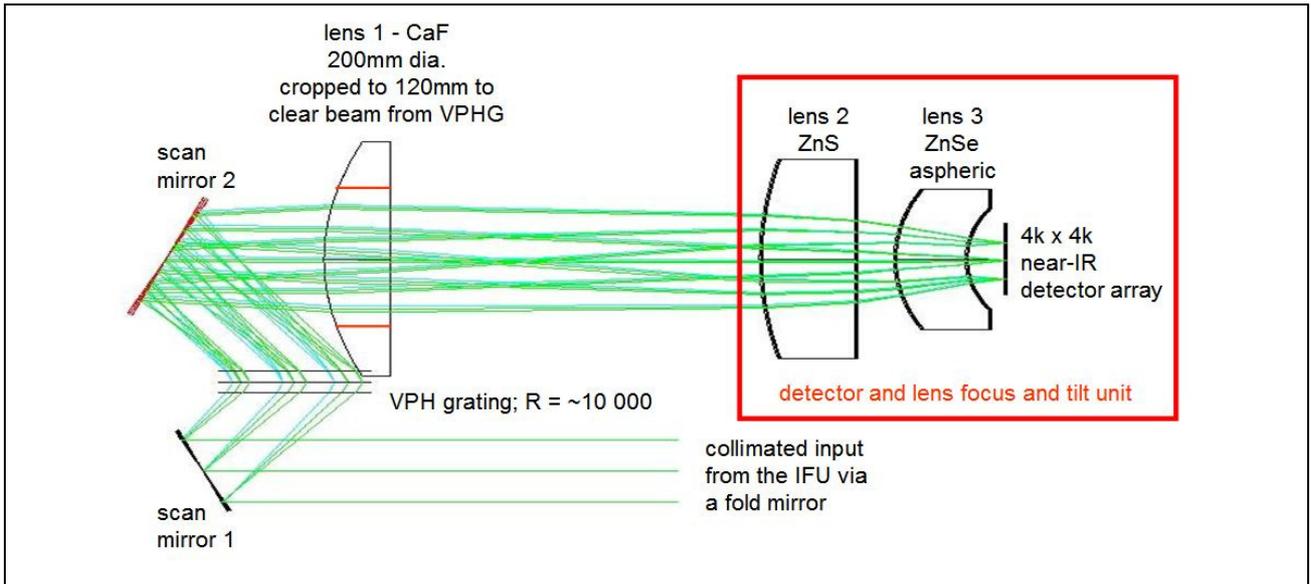

Figure 5: Layout of the Spectrograph optics for a spectral resolution of 10000

Figure 4 and Figure 5 show the Spectrograph configured for normal- and high-resolution observations. In Figure 5 the red lines indicate where L1 has to be trimmed to a rectangular shape (120 x 200 mm) to clear the beam from the VPHG. The high-resolution gratings are extended in the dispersion direction to accommodate the movement of the pupil as the mirrors are scanned.

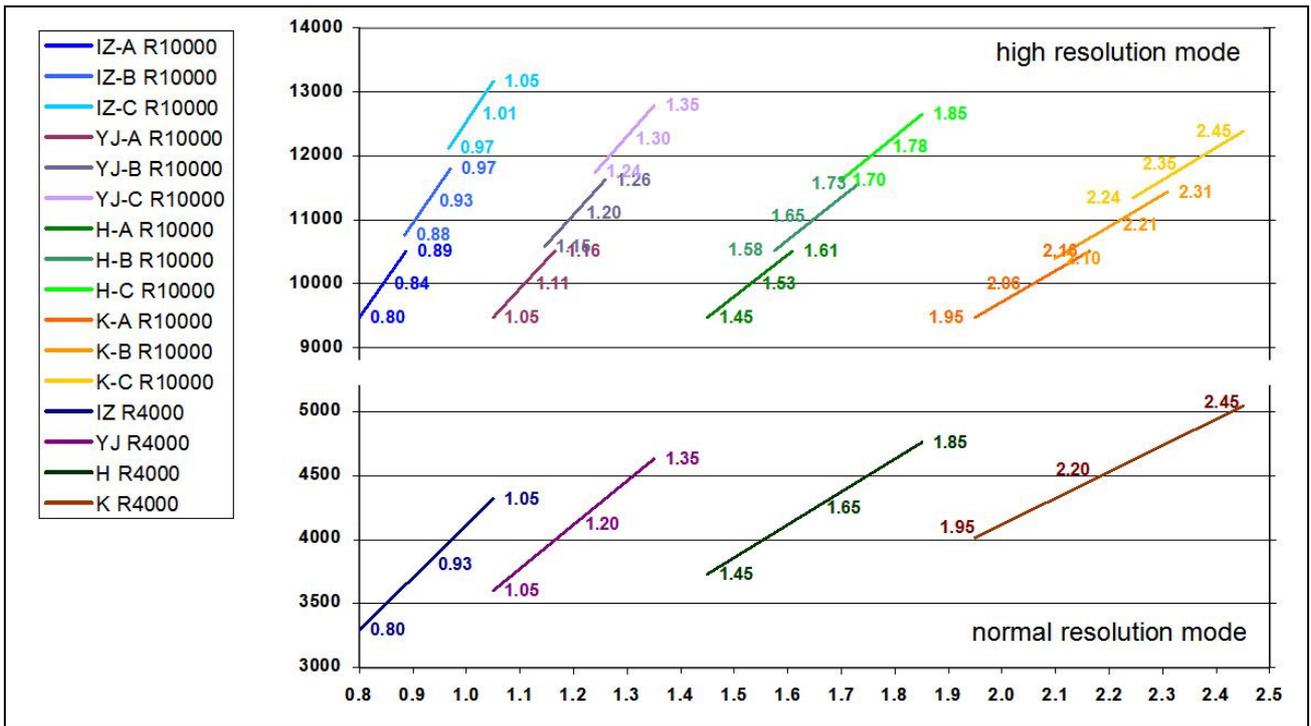

Figure 6: Relationship between wavelength in µm and resolving power for the R=4000 and R=10000 modes

Figure 6 shows the resolving powers and wavebands for the R4K gratings and each of the three settings for the R10K gratings. Observing an entire waveband at R = ~4000 in a single observation uses approximately 2000 pixels on the detector. Curvature of the spectra increases the number of spectral pixels used to ~3000. Aberrations of the current optical design are matched to this illuminated area of detector. For the nominal R10K spectra the grating dispersions were selected to give R=10000 at the mid-point of the first of three equally-spaced sub-bands for each of the four scientific wavebands. This provides some overlap between the three sub-bands. Figure 6 shows that each of the four wavebands can be covered in a single observation at R=4000 and that three observations provide full coverage (with overlap) at R=10000. It is worth noting the scan mirrors allow the start and end points of a given waveband to be adjusted to match specific observations, if needed.

### 3.3 Focus and tilt mechanism

In designing the camera for the Spectrograph a number of options were examined with particular consideration being given to image quality, transmission, cost, packaging and ease of manufacture. For most of these criteria it was found that the introduction of a focus and detector tilt mechanism was beneficial and such a mechanism is therefore part of the baseline design. The image quality (ensquared energy in one pixel) was better for a three lens design with adjustable focus and detector tilt than for an achromatic design with six lenses. The transmission will also be ~6% better (assuming 1% AR coatings) with six fewer glass-air interfaces.

Given that a loss of transmission can never be recovered, except by increasing the length (and cost) of an observation, the decision was taken to increase the number (and cost) of mechanisms to improve the transmission. The experience of the EAGLE Consortium in designing and using cryo-mechanisms made this a low-risk choice. This trade between capital and operating costs, which needs to be reflected in budgeting profiles, is a typical consideration in the design of modern instruments.

The lenses L1 and L2 are similar to lenses in the six-element achromatic design. L3 has an aspheric back surface but the aperture and deviation from a best fit sphere is similar to lens L6 in the camera for KMOS which has been manufactured and is now under test. Lens L1 has a near-parabolic front surface and both L2 and L3 are made from materials which can be diamond turned. The performance of the Spectrograph camera optics is shown in Figure 7 below.

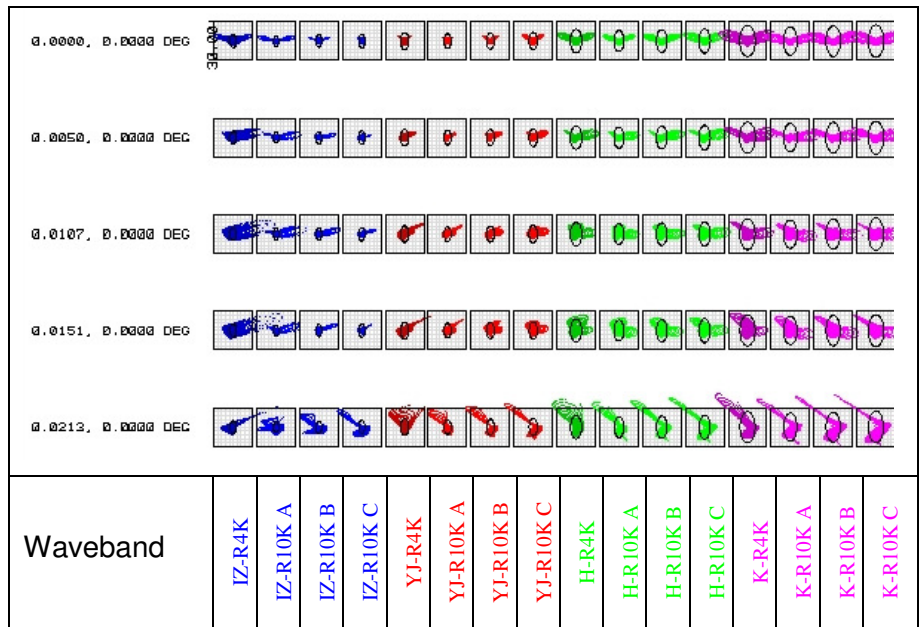

Figure 7: Spot diagrams for the three-element camera

In Figure 7 the spot sizes are shown for the longest wavelength in each passband, which is the worst case. The enclosing box is 2 x 2 pixels and the field locations are at the centre (top) and at 25%, 50%, 75% and 100% of the distance from the centre to the edge of the detector.

Using a focus and tilt mechanism simplifies the AIV (assembly, integration and verification) of the cameras – without them accurate knowledge of the cryogenic refractive indices and detailed modelling of each camera, as-built, are needed to position the detector. Even then, with no means to produce a set of through-focus images, it is hard to determine whether the detector is truly positioned for best image quality. The movements of the focus mechanism for each passband and resolving power are shown in Figure 8.

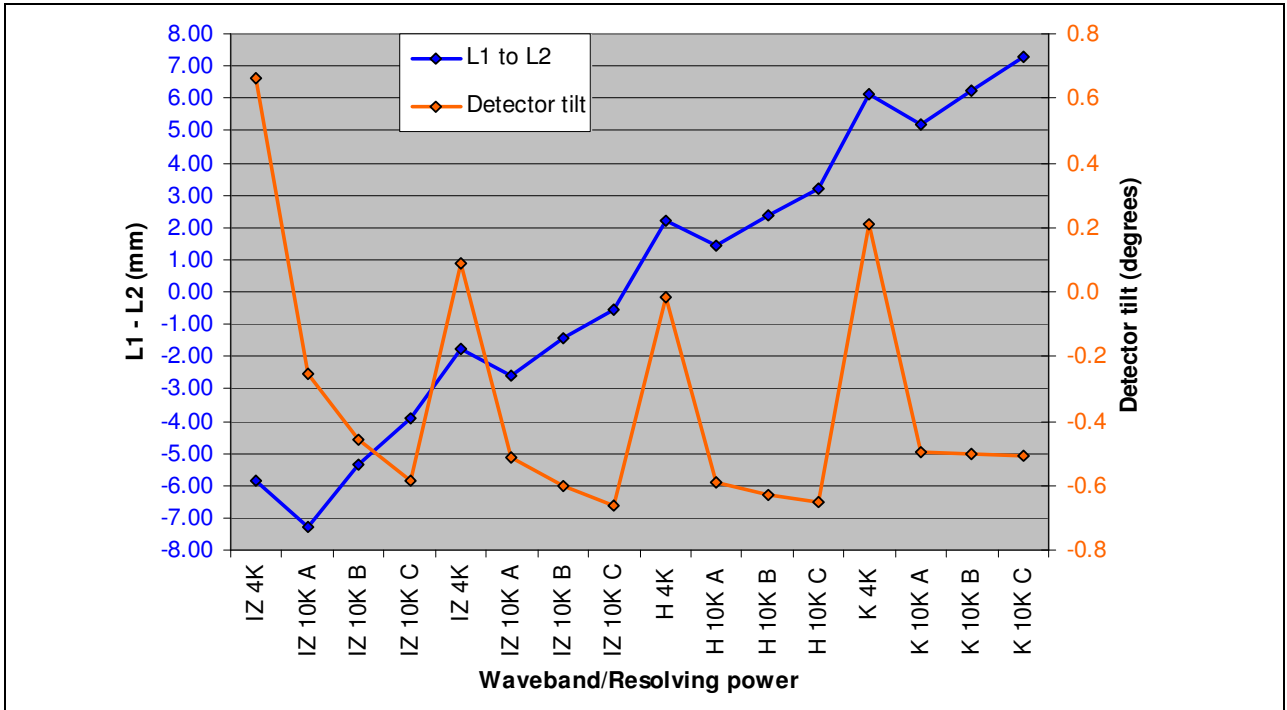

Figure 8: Range of motion for the optics and tilt of the detector for the various passbands and resolutions

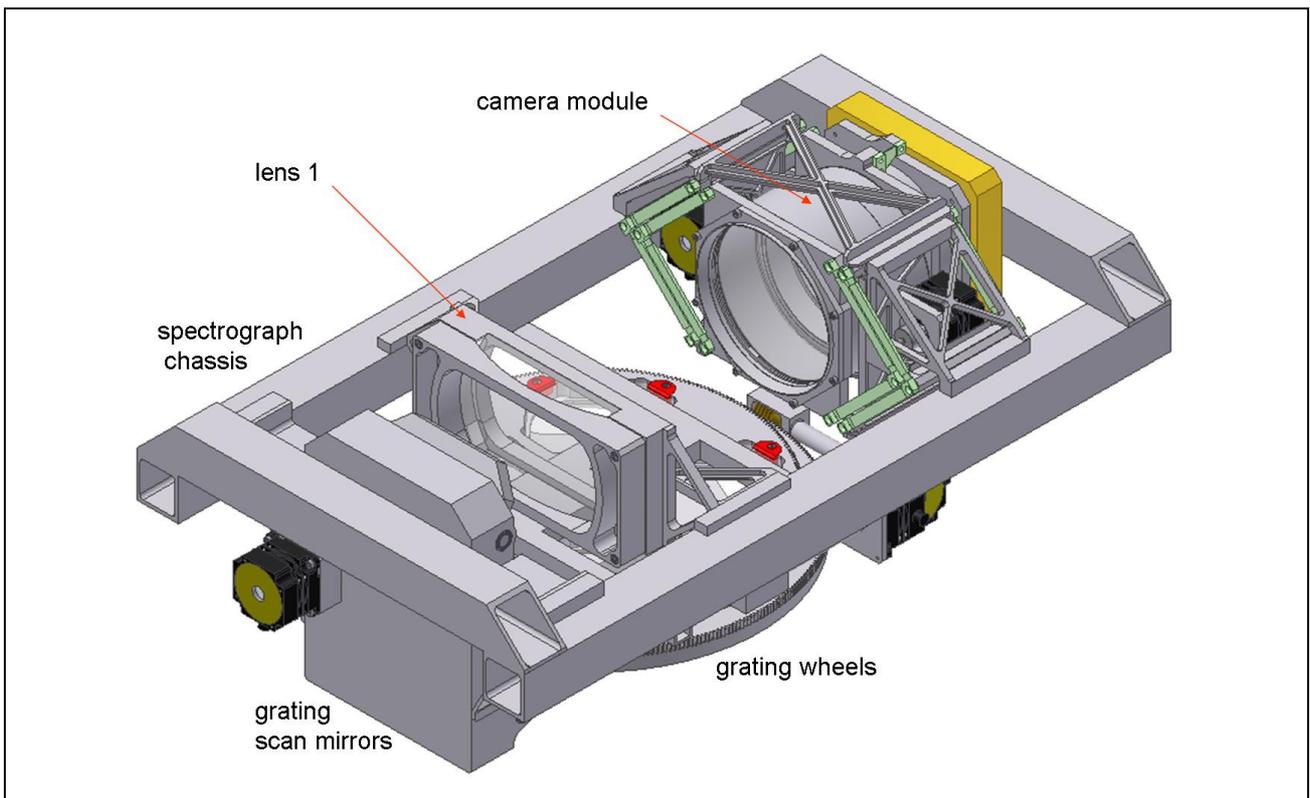

Figure 9: Spectrograph – overall view

Figure 9 shows an overall view of the Spectrograph. The two long rails of the Spectrograph chassis attach to two of the main tubes of the Cryo-structure shown in Figure 10.

## 4. CRYO-MECHANICAL DESIGN

The cryo-mechanical design of the ISS has been influenced by the need for interchangeability, the overall mass budget and the lack of a finally-defined telescope infrastructure. The interchangeability requirements have been addressed by using the 'double-shim' physical interface described above. The mass budget has been met (with margin) by packaging the hardware such that it can be mounted in a cylindrical vacuum vessel – which can be lighter than one with large flat surfaces. Uncertainty about the cryogenic services available at the finished telescope steered the design towards a solution in which the hardware is cooled by an externally-mounted cooling system linked by a number of conductive straps. This allows a change to be made to the type of cooling system without requiring significant re-design of the main opto-mechanics of the ISS. It will also assist in the AIV by allowing testing to be carried out in facilities which have existing cryogenic infrastructure different from that finally used on the E-ELT.

### 4.1 Cryo-structure

The cold structure of the ISS is shown below in Figure 10 in which the larger part of the cylindrical vacuum vessel and the radiation shield have been removed. The aluminium structure consists of two circular end plates separated by three cylindrical columns. One end plate is supported radially and axially by the three support trusses made from glass-epoxy composite. The other end plate is supported (only radially) by three other glass-epoxy trusses. Cross-bracing to improve shear and torsional stiffness will be added as needed and as clearance around the optical path permits. The columns are hollow and act as pre-cool tanks for the structure, IFU and Spectrograph. The detail of the inter-connecting pipes can be altered so that the tanks form part of a recirculating-cryogen cooling system.

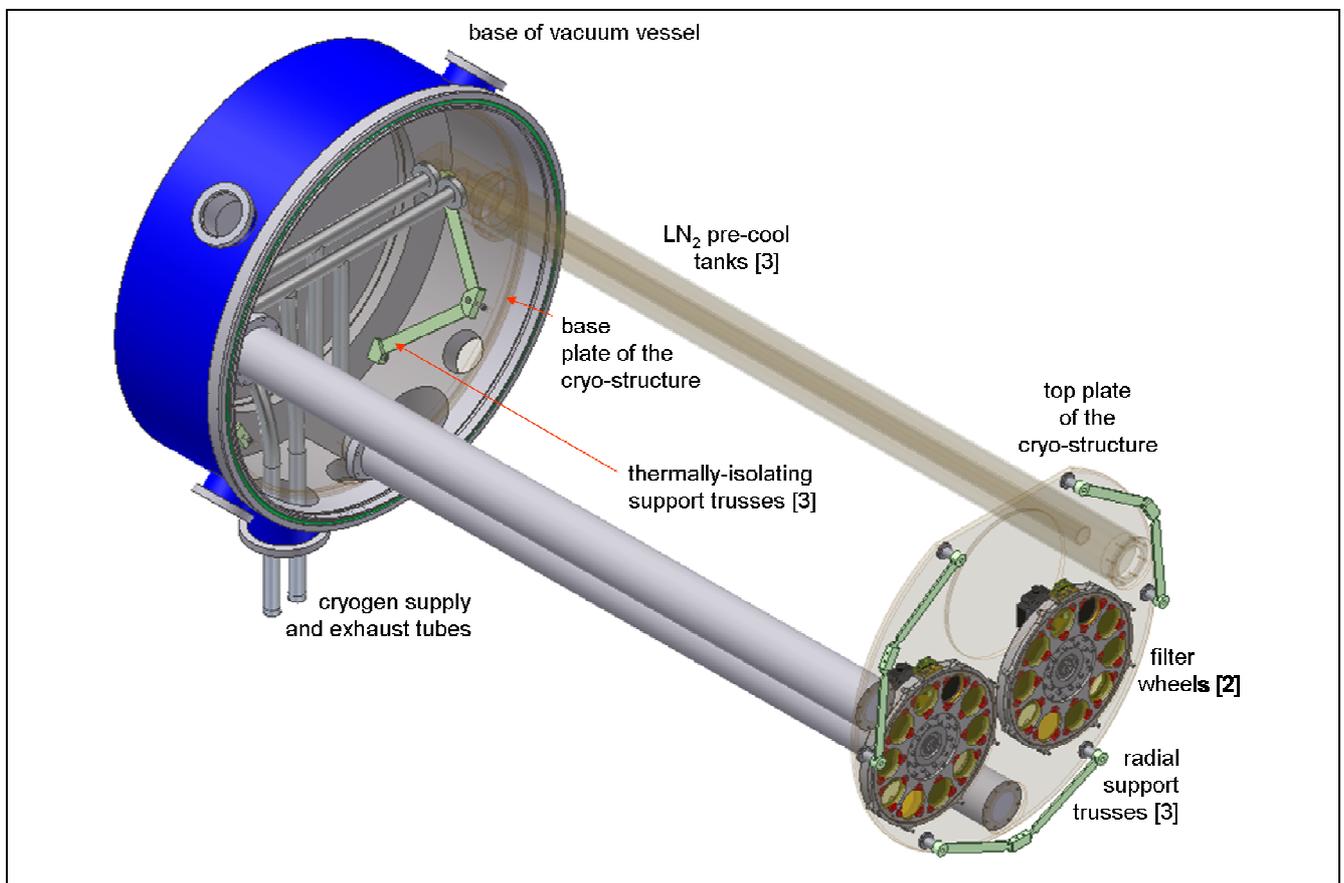

Figure 10: Vacuum Vessel Base and Cryo-structure with radiation shield removed

### 4.2 Cryo-Cooling Module (CCM)

The details of the CCM will depend on:

    The optimum operating temperature of the detector array (to minimise the time needed for an observation)

- The final cryogenic infrastructure of the telescope – closed-cycle, recirculating cryogen, wet Dewar, etc.
- The available cryogenic infrastructure where the AIV takes place.

A place-holder based on a two-stage pulse-tube cooler (shown in Figure 13) was designed as part of the EAGLE Phase A Study. This was chosen in the expectation that the space and mass needed for this would be the greatest of all the various likely solutions. Two nested thermal links pass through a large port in the vacuum vessel base (half-hidden by the nearest pre-cool tank in Figure 10). The outer link attaches to the base plate of the cryo-structure. The inner link passes through an aperture in the cryo-structure base plate and connects flexibly to the detector housing. The CCM shares the vacuum space with the rest of the ISS but can be removed as a unit without disturbing either optics or cryo-mechanisms.

### 4.3 Cryo-Mechanisms

The ISS contains the following cryo-mechanisms :

- two filter wheels (one for each of the science targets) containing passband and order-blocking filters,
- a Scan Mirror Module which directs the beam on to and receives the beam from the VPHG,
- two grating wheels (one for nornal- and one for high-resolution VPHGs),
- a Camera Focus Module which moves the detector, L3 and L2 axially and tilts the detector.

The four wheels are wholly conventional [11] with a cryogenically-prepared stepping motor driving an aluminium wheel with peripheral gear teeth via a Vespel worm. The wheel for the high-resolution gratings has a slot cut in its edge to allow the diffracted beam from the normal-resolution VPHG to pass through. As a result it cannot rotate continuously and is treated for motor control and software design as a linear mechanism rather than a rotary one.

### 4.4 Scan Mirror Module (SMM)

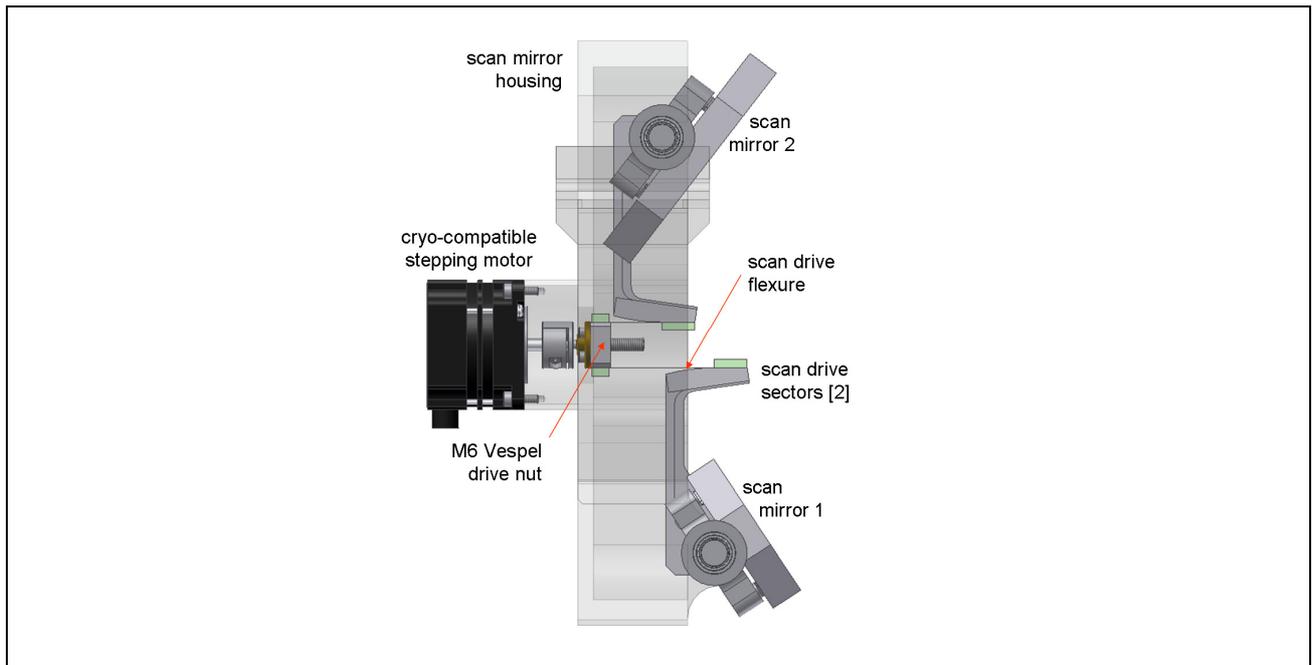

Figure 11: Ghosted diagram of the Scan Mirror Module showing the flexure drive

The SMM shown in Figure 11 uses a novel cryo-mechanism in which two plane mirrors are driven through the same angle in opposite directions by a single leadscrew and nut. The drive is transmitted by a glass-epoxy band (TBC) which wraps around the periphery of sectors attached to the mirrors. The sectors need to be pre-loaded to keep the bands in tension and to remove backlash from the system. One motor step equates to a change of 12 micro-radians in the angle of each mirror. One of the mirror/sector sub-assemblies is shimmed to set the initial alignment of the pair of mirrors and the whole module is attached to the Spectrograph chassis at three points using a 'double shim' interface.

### 4.5 Camera Focus Module (CFM)

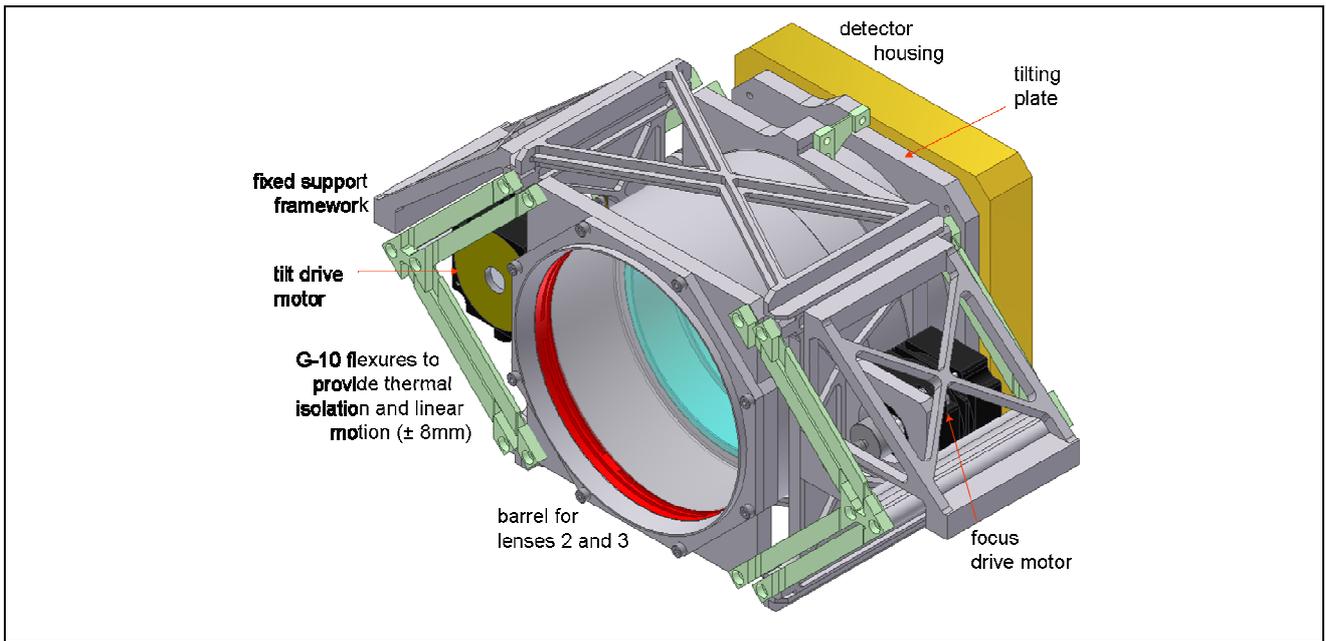

Figure 12: Camera Focus Module and Detector Housing

The CFM shown in Figure 12 consists of an aluminium lens barrel, containing lenses L2 and L3, which is suspended on four double-parallel flexures made from glass-epoxy composite. A tilting plate is attached to one end of the barrel by two triangular blade flexures. The detector housing is attached to this plate and thus moves axially with the lens barrel. The barrel carries two cryogenically-prepared stepping motors. One drives the barrel axially via a leadscrew and static nut. The other tilts the plate, and thus the detector, again using a leadscrew and nut. A triangulated aluminium support framework encloses the barrel and attaches to the Spectrograph chassis by 'double-shim' interfaces.

## 5. CONCLUSIONS

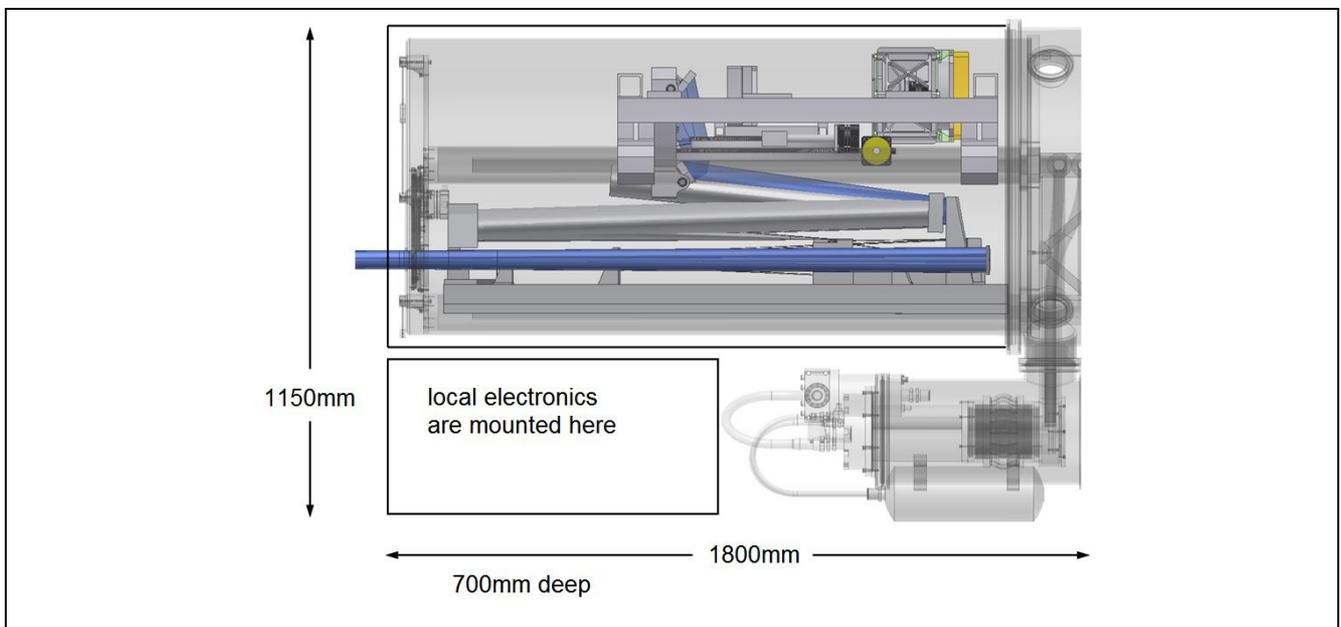

Figure 13: Ghosted view of the ISS showing overall layout and principal dimensions

At the end of the EAGLE Phase A Instrument Study, the ISS illustrated in Figure 13 meets functional requirements for spatial and spectral sampling of the two 1.65 x 1.65 arcsec AO-corrected science fields delivered to it. Initial analysis shows that it will fit within the allocated volume and that the required movements can be achieved using standard designs and practice for cryogenic mechanisms. Optical analysis shows that for a large fraction of the field and spectral points the IFU and Spectrograph have ensquared energy (2 x 2 pixels) > 0.95.

A design using a prudent combination of innovative and conventional technologies has allowed two near-IR integral-field spectrographs to be fitted in a volume of $1.5m^3$ and weigh not more than 500kg. While this might be considered moderately large by current standards, the ISS has been designed for a 42m telescope and has a resolution-luminosity product which is twice that of a comparable multi-object integral-field near-IR spectrograph for an 8 metre telescope. As currently envisaged, EAGLE has ten identical ISSs.